\documentstyle[aps,pre,psfig,epsfig]{revtex}
\newcommand \beq{\begin{eqnarray}}
\newcommand \eeq{\end{eqnarray}}
\newcommand \be{\begin{eqnarray}}
\newcommand \ee{\end{eqnarray}}
\newcommand{\set}[2]{\newcommand{#1}{#2}}
\set{\pa}{\partial \over \partial\, }
\set{\leftvector}{\stackrel{\leftarrow}{\partial }}
\set{\rightvector}{\stackrel{\rightarrow}{\partial }}
\begin{document}
\twocolumn[\hsize\textwidth\columnwidth\hsize
           \csname @twocolumnfalse\endcsname
\title{Exactly solvable model of three interacting particles in an
 external magnetic field} 
\author{E.\ P.\ Nakhmedov$^{1,2,6}$ \and K.\ Morawetz$^{1}$
 \and M.\ Ameduri$^{3,4}$ \and A.\ Yurtsever$^5$ \and C. Radehaus$^6$}  
\address{$^1$Max Planck Institute for the Physics of Complex Systems,
 N\"othnitzer Str.\ 38, 01187 Dresden, Germany\\
 $^2$Azerbaijan Academy of Sciences, Institute of Physics, H.\ Cavid 33,
 370143 Baku, Azerbaijan\\
 $^3$Laboratory of Elementary Particle Physics, Cornell University,
 Ithaca, NY 14853, USA\\
 $^4$Weill Cornell Medical College in Qatar, Qatar Foundation,
 Doha, Qatar\\
 $^5$Dept.\ of Mathematics, Fatih University, B\" uy\" uk\c cekmece,
 Istanbul, Turkey\\
$^6$Technische Universit\"at Chemnitz, D-09107 Chemnitz}
\maketitle
\date{\today}
\begin{abstract}
The quantum mechanical problem of three identical particles, 
moving in a plane and interacting pairwise via a spring potential, 
is solved exactly in the presence of a magnetic field. Calculations of the
pair--correlation function, mean distance and the cluster area show a 
quantization of these parameters. Especially the pair-correlation function 
exhibits a certain number of maxima given by a quantum number. 
We obtain Jastrow pre-factors which lead to 
an exchange correlation hole of liquid type, even in the presence of the attractive interaction 
between the identical electrons. 
\end{abstract}
\pacs{PACS numbers: 05.30.-d, 05.20.Dd, 05.60.+w, 72.20.Ht  }
\vskip2pc]
\section{Introduction}

The Laughlin wave function \cite{L83,L87} was introduced to describe
the ground state of a two-dimensional (2D) long-range correlated electron
gas in the fractional quantum Hall regime. The strong short-distance
repulsion between any two particles is described by the Jastrow prefactor
in the wave function. The same Jastrow prefactor appears also in the exact
wave functions for $N$ particles interacting pairwise through a potential
of the form $1/(x_{i}-x_{j})^{2}$, first obtained in one dimension by
Calogero \cite{C69,Ha94}, and later extended to two and higher
dimensions \cite{CM73,KR97,K98}. These wave functions lead to the
Wigner-Dyson distribution function \cite{SLA94} for $N$ particles.

A natural question therefore is whether the origin of this Jastrow
prefactor is due to special features of the interactions mentioned above,
or it is a generic property of models with short-range interactions.

The recent advances in the theory of disordered systems show that
the distribution function of physical parameters of the mesoscopic system,
such as the density of states (DOS) or the conductance, is represented
by the Wigner-Dyson function in the weak localization regime (see,
e.g., \cite{ALW91} for a review). Physically, this describes a
repulsion between the energy levels. Averaging the partition function
of a disordered system over the Gaussian-distributed one-particle impurity
potentials results in an effective Hamiltonian with short-range
interactions. This clearly shows that level repulsion is not a property
of just few special interaction potentials.

In a recent paper \cite{NM01}, a new mechanism for the formation of
three-particle clusters in Si-MOSFET structures and GaAs/AlGaAs
heterojunctions has been studied. The exchange type interactions between
$2D$ band electrons in the inversion layer and charged impurities in
the oxide of the MOSFET can lead to an effective three-particle attractive
interaction. A qualitative analysis shows that the ground state of the
three particles becomes unstable with respect to arbitrary small 
attractive interactions between them. An attraction between three
particles leads to the formation of bound states with negative
energies \cite{NM01}. Although a weak attractive three-particle interaction
does not produce a transition to a liquid state of the electron gas as
described in Laughlin's approach, the ground state with three-particle
clustering could be energetically favored in the fractional quantum
Hall regime. This is a consequence of the occurrence of an effective
pairing \cite{NM01}, which further reduces the ground state energy. 

It is interesting to study the internal structure of a particular cluster
with three particles, in order to find out if its wave function contains
a Jastrow prefactor. This would prevent the particles from approaching
each other even in the attractive case. In this paper we study this
question by modeling a generic three-particle attractive potential by a
pairwise spring-like interaction. We show that this choice of potential
allows for an exact solution even in the presence of a constant external
magnetic field.

Exact results obtained for solvable models are important to understand
general features of three-particle problems and to test approximate
solutions, such as the ones obtained from the widely used Fadeev's
equations \cite{F65}.
Very few exact solutions of three-particle problems have been found,
mostly for one-dimensional \cite{Dodd70,HT74,MB78} or for three-dimensional
\cite{FGP80,BM92,BM94} spinless particles. In this paper we present an
exact solution for a model of particles with spin.

The topology of the three-particle cluster may be either
of a string form, when two particles with opposite spin are placed
in the same spatial point, or of a triangle form when all three particles
lie far from each others. In the former case the number of the 
spin configurations is equal to two (\emph{doublet states}) with
total spin $S=1/2$. Apart from this \emph{doublet states}, a triangular
clustering of \emph{quartet states} may be also realized. This would
correspond to four different symmetric spin wave functions with a total spin
$S=3/2$. In a strong magnetic field this quartet states are the only
relevant ones.


\section{The model}
The Hamiltonian of the model is 
\be
\hat{H}&=& -\frac{\hbar^2}{2m^*}\sum_{\alpha = 1}^3
\Big(\frac{\partial}{\partial {\bf r}_{\alpha}} - i\frac{e}{\hbar
  c}{\bf A}_{\alpha}\Big)^2 
\nonumber\\&&
+ \frac{\kappa}{2}[({\bf r}_1- {\bf
  r}_2)^2 + ({\bf r}_1- {\bf r}_3)^2 + ({\bf r}_2- {\bf  r}_3)^2] ,
\label{H}
\ee
where ${\bf r}_{\alpha}$ ($\alpha = 1,2,3$) are the two-dimensional
($2D$) position vectors of electrons with effective mass $m^*$. 
The spring strength is chosen as $\kappa = m^*\omega_0^2 /3$
where $\omega_0$
is the specific frequency of the electron's relative vibration. ${\bf
  A}_{\alpha}= \frac{B}{2}\{-y, x,0\}$ is the symmetric-gauge 
vector potential at the position of the 
$\alpha$-th particle for a magnetic field ${\bf B}=\{0,0,B/2\}$
perpendicular to the electronic inversion layer.

In order to diagonalize the Hamiltonian we introduce the
new coordinates $\{{\bf r}, {\bf \xi}, {\bf \eta}\}$,
\begin{eqnarray}
{\bf r} &=& \frac{1}{\sqrt 3}({\bf r}_1 + {\bf r}_2 + {\bf r}_3), \nonumber\\
{\bf \xi} &=&\frac{1}{\sqrt 2} ({\bf r}_1 - {\bf r}_2), \nonumber\\
{\bf \eta} &=& \sqrt{\frac{2}{3}} \left (\frac{{\bf r}_1 + {\bf
    r}_2}{2} -{\bf r}_3 \right )  ,
\label{coor}
\end{eqnarray}
where ${\bf r}$ is the center-of-mass coordinate, ${\bf \xi}$ is the relative
coordinate of the particles 1 and 2 and ${\bf \eta}$ is the relative
coordinate of the third particle with respect to the center-of-mass of the 
particles 1 and 2. In the new coordinates, the Hamiltonian is decoupled,
\begin{equation}
\hat{H} = \hat{H_{{\bf r}}} + \hat{H_{{\bf \xi}}} +\hat{H_{{\bf \eta}}} ,
\end{equation}
where
\begin{eqnarray}
\hat{H_{{\bf r}}}&=& -\frac{\hbar ^2}{2m^*}\frac{\partial ^2}{\partial
  {\bf r}^2} + \frac{3\hbar^2}{8m^*l_B^4}{\bf r}^2
+i\frac{\hbar^2}{2m^*l_B^2}({\bf r}\times \nabla_{\bf r})_z ,
\nonumber\\&& \label{Hr}\\
\hat{H_{{\bf \xi}}}&=&-\frac{\hbar ^2}{2m^*}\frac{\partial ^2}{\partial
  {\bf \xi}^2} + \big(\frac{\hbar^2}{8m^*l_B^4}+
\frac{3\kappa}{2}\big){\bf \xi}^2 
+i\frac{\hbar^2}{2m^*l_B^2}({\bf \xi}\times \nabla_{\bf \xi})_z ,
\nonumber\\&&\\
\label{Hxi}
\hat{H_{{\bf \eta}}}&=&-\frac{\hbar ^2}{2m^*}\frac{\partial ^2}{\partial
  {\bf \eta}^2} + \big(\frac{\hbar^2}{8m^*l_B^4}+
\frac{3\kappa}{2}\big){\bf \eta}^2 
+i\frac{\hbar^2}{2m^*l_B^2}({\bf \eta}\times \nabla_{\bf \eta})_z.
\nonumber\\&&\label{Heta}
\end{eqnarray}
The last terms in the Hamiltonians given by
Eqs.\ (\ref{Hr})-(\ref{Heta}) are proportional to the angular 
momentum and $l_B^2= \frac{\hbar c}{e B}$ is the magnetic length.

The wave function $\Psi$  of the initial Hamiltonian, Eq.\ (\ref{H}),
can be presented as the product of three wave functions $\psi$ which are
eigenfunctions of Eqs.\ (\ref{Hr})-(\ref{Heta})
\begin{equation}
\Psi ({\bf r}_1,{\bf r}_2,{\bf r}_3)= \psi (r)\psi (\xi) \psi (\eta).
\label{Psi}
\end{equation} 
Schr\"odinger's equation $\hat{H}\psi = E\psi$ with the Hamiltonian given by
one of the Eqs.\ (\ref{Hr})-(\ref{Heta}) is easily solved in a polar
coordinate system where, e.g., Eq.\ (\ref{Hxi}) is written as
\be
\hat{H}_{\rho}&=&-\frac{\hbar ^2}{2m^*}\Big(\frac{\partial ^2}{\partial
  \rho^2}+\frac{1}{\rho}\frac{\partial}{\partial \rho}
+\frac{1}{\rho^2}\frac{\partial^2}{\partial \varphi ^2}\Big)   
-i\frac{\hbar^2}{2m^*l_B^2}\frac{\partial}{\partial \varphi}
\nonumber\\&&+ 
\Big(\frac{\hbar^2}{8m^*l_B^4}+
\frac{3\kappa}{2}\Big)\rho^2
\label{Hrho}
\ee
and $\rho$ and $\varphi$ are the radial and angular variables
in the polar coordinate systems.

The Hamiltonian (\ref{Hrho}) commutes with the angular momentum operator
$\hat{L}_z = -i\hbar \frac{\partial}{\partial \varphi}$,
$[\hat{H_{\rho}},\hat{L}_z]=0$. Therefore the wave function
$\psi(\rho,\varphi)$ of the Hamiltonian $\hat{H}_{\rho}$ can be chosen
to be of the form
\begin{equation}
\psi_m(\rho,\varphi) = e^{im\varphi}R(\rho),
\label{wave1}
\end{equation} 
where $m$ is the angular momentum for a $2D$ system and $m= \pm 1,\pm 2,\pm 3
\dots$.

The asymptotic behavior of the radial wave function $R(\rho)$ can be
separated as
\begin{equation}
\bar{R}(z) = z^{\frac{|m|}{2}}e^{-\frac{1}{2}z}f(z),
\end{equation} 
where $z=
\frac{\rho^2}{2l_B^2}\sqrt{1 +\frac{4\omega_0^2}{\omega_B^2}} $ is a
dimensionless coordinate, and $\omega_B = \frac{e B}{m^*  c}$ is the
cyclotron frequency.

The function $f(z)$ satisfies the confluent hypergeometric equation
\begin{equation}
f(z) = F\Big\{-\big(\frac{\beta}{\sqrt{1+4\omega_0^2/\omega_B^2}}
  -\frac{1}{2}(|m|+1)\big), 1+|m|,z \Big\}
\end{equation} 
where $\beta =
\frac{E}{4\omega_B} -\frac{m}{2}$ is the dimensionless energy spectrum. 
Convergence requires the first coefficient of the confluent hypergeometric
function to be a negative integer. The function then reduces to
the generalized Laguerre polynomials and gives the following
energy spectrum
\begin{equation}
E_{n_i,m_i}= \hbar \omega_B \sqrt{1+\frac{4\omega_0^2}{\omega_B^2}}(n_i +
\frac{|m_i|+1}{2} ) + \frac{\hbar \omega_B}{2} m_i ,
\end{equation}
where $i=1,2,3$ denotes the number of non-interacting quasi-particles
introduced by means of a normal coordinate transformation.
The total energy of a three- particle cluster is expressed as a sum of the
energies of three non-interacting quasi-particles. Redefining the initial
quantum numbers as $n_0= n_1 + \frac{1}{2}(|m_1|+m_1)$, $n= 2(n_2 +
n_3) +|m_2| +|m_3|$ and $m= m_2 + m_3$, the total energy becomes
dependent on three exact quantum numbers,
\be
E(n_0,n,m) &=& \hbar \omega_B (n_0 + \frac{1}{2}) 
\nonumber\\&&
+ \frac{\hbar
  \omega_B}{2} \sqrt{1+ \frac{4\omega_0^2}{\omega_B^2}}(n+2)
+\frac{\hbar \omega_B}{2} m.
\label{E}
\ee
The case corresponding to three free particles in an external magnetic
field is given by the two exact quantum numbers $n =
\sum^3_{i=1} (n_i + |m_i|/2)$ and $m = \sum^3_{i=1} m_i$. 
The normalized wave function of the $i$-th quasi-particle can be written in
the form 
\be
&&\psi_{n_i,m_i}(\rho,\varphi )= \frac{1}{\sqrt{\pi}}
\left ({\sqrt{1+\frac{4\omega_0^2}{\omega_B^2}}\over 2 l_B^2}\right )^{\frac{|m_i|+1}{2}}
\Big[\frac{n!}{(n_i+|m_i|)!}\Big]^{1/2}
\nonumber\\
&&\times
e^{im_i\varphi} \rho^{|m_i|}
{\rm e}^{-\sqrt{1+\frac{4\omega_0^2}{\omega_B^2}}
\frac{\rho^2}{4l_B^2}} L^{|m_i|}_{n_i}\left (\sqrt{1+
  \frac{4\omega_0^2}{\omega_B^2}} \frac{\rho^2}{2l_B^2}\right ). 
\label{wave}
\ee
In order to construct the total wave function for a three particles cluster,
the spatial
wave functions corresponding to Eqs.\ (\ref{Hr})-(\ref{Heta}) must be
multiplied by the appropriate spin wave functions and the final
expression must be antisymmetric.

\section{Wave functions and correlators}

All spin wave functions of the quartet states $|S={3\over 2},s_z \rangle$,
corresponding to the total spin number $S=3/2$ and its $z$-component
$s_z$, are symmetric
\begin{eqnarray}
&&|3/2,3/2\rangle  = \alpha_1 \alpha_2 \alpha_3\\
&&|3/2,1/2\rangle  = \frac{1}{\sqrt 3}\{\alpha_1 \alpha_2 \beta_3 +
\alpha_1 \beta_2 \alpha_3 + \beta_1 \alpha_2 \alpha_3 \} \\
&&|3/2,-1/2\rangle = \frac{1}{\sqrt 3}\{\alpha_1 \beta_2 \beta_3 +
\beta_1 \beta_2 \alpha_3 + \beta_1 \alpha_2 \beta_3  \} \\
&&|3/2,-3/2\rangle = \beta_1 \beta_2 \beta_3
\end{eqnarray}  
where $\alpha_i$ and $\beta_i$ are the spinors of the $i$-th electron
corresponding to spin up and down, respectively. 
Therefore, the antisymmetric orbital part of the wave function reads
\begin{eqnarray}
&&\Psi^Q({\bf r_1},{\bf r_2},{\bf r_3})=C\Psi_{n_1,m_1}(r;\omega_0=0)
\nonumber\\
&&\times \{
\Psi_{n_2,m_2}(\xi)\Psi_{n_3,m_3}(\eta) 
-
\Psi_{n_2,m_2}(-\xi)\Psi_{n_3,m_3}(\eta)
\nonumber\\
&&
+\Psi_{n_2,m_2}(\tilde\xi)\Psi_{n_3,m_3}(\tilde\eta)
-\Psi_{n_2,m_2}(-\tilde \xi)\Psi_{n_3,m_3}(\tilde\eta)
\nonumber\\
&&
+\Psi_{n_2,m_2}(\tilde{\tilde \xi})\Psi_{n_3,m_3}(\tilde{\tilde\eta})
-\Psi_{n_2,m_2}(-\tilde{\tilde \xi})\Psi_{n_3,m_3}(\tilde{\tilde\eta})
 \} ,
\label{PsiA}
\end{eqnarray}
where $C$ is a normalization constant and we have used
the notation
\be
\tilde{\bf \xi}&=&\frac{{\bf r_2}-{\bf r_3}}{\sqrt 2}
={\sqrt{3}\over 2} {\bf \eta} -{{\bf \xi}\over 2} ,
\nonumber\\
\tilde {\bf \eta}&=&\sqrt{2\over 3} \big(\frac{{\bf
r_2}+{\bf r_3}}{2}- 
{\bf r_1}\big)
=-{{\bf \eta}\over 2}-{\sqrt{3}\over 2} {\bf \xi} ,
\nonumber\\
\tilde{\tilde {\bf \xi}}&=& \frac{{\bf r_3}-{\bf r_1}}{\sqrt 2}
=-{\sqrt{3}\over 2} {\bf \eta} -{{\bf \xi}\over 2} ,
\nonumber\\
\tilde{\tilde {\bf \eta}}&=&\sqrt{2\over 3}\big(\frac{{\bf
r_1}+{\bf r_3}}{2}- {\bf r_2}\big)=-{{\bf \eta}\over 2}+{\sqrt{3}\over 2} {\bf \xi}.
\label{not}
\ee

\begin{figure}
\psfig{file=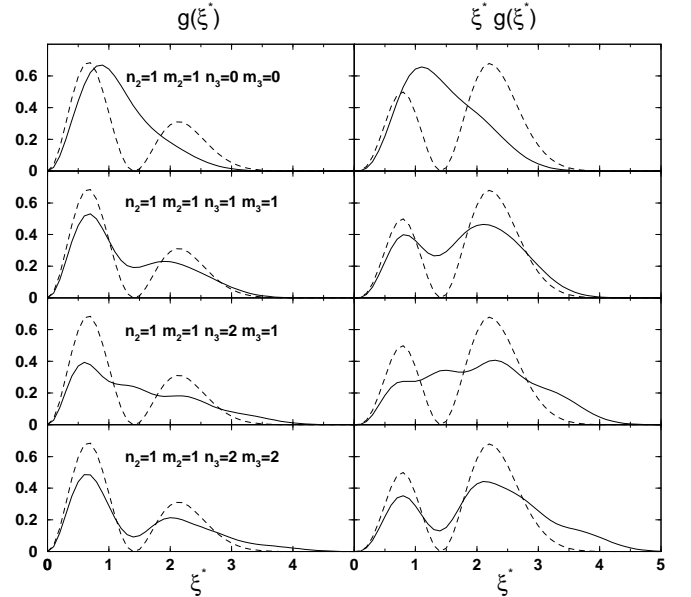,width=8cm,angle=-90}
\caption{The normalized pair correlation function $g(\xi ^*)$
versus the dimensionless distance between two particles $\xi ^*$
for $n_2=1$ and different values of the quantum numbers $n_3, m_3$. 
The dotted lines are the pair correlation function only from the diagonal elements 
of $|\Psi|^2$ according to (\protect\ref{paird}).
\label{dat1}
}
\end{figure}

Notice that the inversion of the coordinates in polar
basis leads to $\rho \to \rho$ and $\varphi \to \varphi + \pi$.
Under this transformation, the wave function acquires a factor $(-1)^m$,
which defines its parity.
It is easy to see that $\Psi^A({\bf r_1},{\bf r_2},{\bf r_3}) =0$
for $m_2$= even integer. The privileged dependence of the wave
function on $m_2$ is originated by the asymmetry in the new coordinates
(\ref{coor}). For $m_2$ = odd integer, we get
\begin{eqnarray}
\Psi^Q({\bf r_1},{\bf r_2},{\bf r_3})&=& C \Psi_{n_1,m_1}({\bf r};\omega_o=0)
\nonumber\\
&\times& 
\left \{
\Psi_{n_2,m_2}(\xi)\Psi_{n_3,m_3}(\eta)
\right .
\nonumber\\&&
+\Psi_{n_2,m_2}(\tilde \xi)\Psi_{n_3,m_3}(\tilde \eta)
\nonumber\\&&
\left . 
+\Psi_{n_2,m_2}(\tilde {\tilde \xi})\Psi_{n_3,m_3}(\tilde {\tilde \eta})
\right \}.
\label{Psianti}
\end{eqnarray}

In high magnetic fields the polarized spin structure is
characterized by this \emph{quartet states} since all states are aligned.  
The valuable 
feature of the spin-polarized quartet states is that they permit only 
odd values of $m_2$ and exclude even ones, which is the situation 
observed in the fractional quantum Hall experiments.
The wave function $\Psi^Q$ given by
Eq.\ (\ref{Psianti}) has the property that $\Psi^Q({\bf r_1},{\bf
  r_2},{\bf r_3}) \vert _{{\bf r_i}\to {\bf r_j}} = 0$ for arbitrary
chosen $i \ne j =1,2,3$ . This corresponds to the behavior of the Jastrow 
prefactor $f({\bf r_i} - {\bf r_j})$ in Laughlin's 
ground state wave function \cite{L83,L87}.

\begin{figure}
\psfig{file=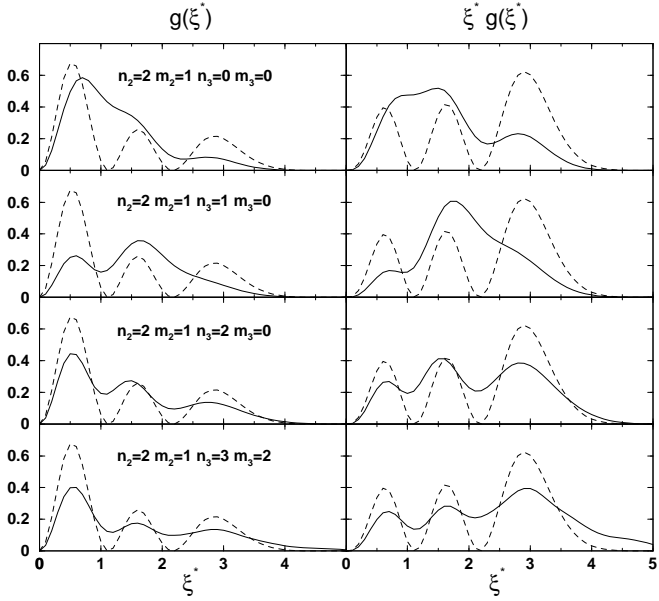,width=8cm,angle=-90}
\caption{The pair correlation function $g(\xi ^*)$ versus $\xi^*$ for $n_2=2$ and different values of
$n_3$ and $m_3$.
\label{dat2}
}
\end{figure}

\begin{figure}
\psfig{file=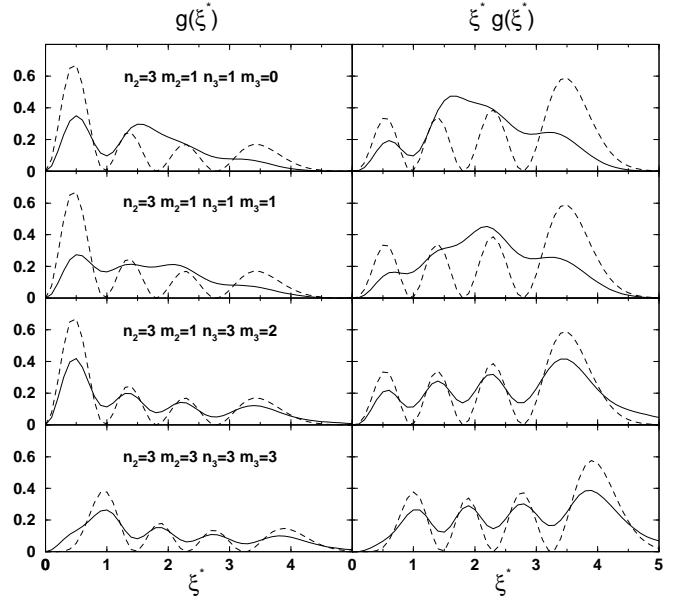,width=8cm,angle=-90}
\caption{The dependence of $g(\xi^*)$ on space coordinate for $n_2=3$ and different values of $n_3$ and $m_3$.
\label{dat3}
}
\end{figure}

In order to get more insight into the wave function we plot in figures \ref{dat1}-\ref{dat3} the (radial) pair correlation function
\be
G(|\xi|)&=&\int\limits_0^{2\pi} d \phi_\xi \int d^{2}r \int d^{2} \eta 
\left |\Psi^Q\right |^2
\nonumber\\
&=&{\sqrt
{1+\frac{4\omega_0^2}{\omega_B^2}} \over l_B^2
}
g(|\xi^*|),
\label{pair}
\ee
where the dimensionless coordinate $\xi^{*}$ is
\be
\xi^*={\left (1+\frac{4\omega_o^2}{\omega_B^2}\right )^{1/4}\over \sqrt{2} l_B}\xi.
\label{diff}
\ee
This pair correlation function gives the probability of finding
two particles at a given distance in the presence of a third particle.
We find that, owing to the presence of the third
particle, a shell structure appears with certain maxima in the probability
at special distances. 
The number of these maxima (shells) is defined by the quantum number $n_2$ 
and is equal to $n_2 + 1$.
Therefore, the quantum number $n_2$ can be interpreted as the main
characteristic of the number of resonant states or the number of nearest
neighbors if one considers this as the onset of a liquid like 
behavior. 

This interpretation is supported by the following reasoning. The pair
correlation function starts at zero since the antisymmetrization due to
Pauli-blocking induces the exchange correlation hole. 
A small $\xi$ expansion then yields [appendix~\ref{abc}]
\be
g(|\xi^*|)&=&a |\xi^*|^{2 m_2}+(b+c |\xi^*|^{2 m_2})|\xi^*|^2 +o(|\xi^*|^3)\nonumber\\
&=&\left \{\begin{array}{ll}(a+b)|\xi^*|^2 ,& m_2=1\cr b |\xi^*|^2 +
a |\xi^*|^{2 m_2} ,& m_2>1\end{array}\right .+o(|\xi^*|^3) 
\label{abc1}
\ee
since $m_2$ is odd. The regime of $\xi^* <1$ can be realized, according to Eq.(\ref{diff}),
for an arbitrary finite distance between particles by sufficiently reducing both $\omega_0$ 
and $\omega_H$. 

The characteristic property of the Laughlin wave function is that $g(z)$ goes
to zero as $z^{2 m}$, whereas in the case of the Wigner crystal the pair--correlation 
function goes to zero as
$1-\exp{(-a z^2)}\propto z^2$. This was the reason for 
the superiority of the variational Laughlin wave function. In our direct computation,
starting from the expression of antisymmetric wave function $\Psi^Q$, we obtain 
the Wigner crystal behavior. However, the Laughlin exchange correlation hole
becomes transparent if we use only the first diagonal term in
$|\Psi^Q|^2\propto C^2 |\Psi_{n_1,m_1}(r;\omega_0=0)|^2
|\Psi_{n_2,m_2}(\xi)|^2|\Psi_{n_3,m_3}(\eta)|^2$ which is the first term in (\ref{int}). 
We obtain the analytical result
\be
g^{\rm diag1}\left (\left |\xi^*\right |\right )&=&{
n_2! \over (n_2+|m_2|)!}\xi^{*2 |m_2|}\left (L_{n_2}^{|m_2|}(\xi^{*2})\right )^2 
  {\rm e}^{-\xi^{*2}}.
\label{paird}
\ee
Representing the Laguerre polynomials as
$L^m_n(x)= \sum_{k=0}^n \left (\begin{array}{l}n+m \cr n-k\end{array}\right ) 
\frac{(-x)^k}{k!}$, the following asymptotic behavior of
$g^{{\rm diag}}$ is obtained 
\begin{equation} 
g^{{\rm diag}}(|\xi^*|) ={(n_2+|m_2|)!\over n_2! m_2!^2 }\xi^{*2 |m_2|}\left (1+o(\xi^2)\right ).
\end{equation} 
We see that the typical $z^{2 |m_2|}$ behavior of Laughlin states appears.
Therefore, we consider the quantum number $|m_2|$ as the filling factor in 
analogy to Laughlin's exchange correlation hole  behavior.

The approximation leading to this behavior has consisted in neglecting
certain crossed terms. 
Here we should note that when considering the paircorrelation function we 
design in principle two out of the three particles differently than the 
third particle. In this case we don't need to antisymmetrize the wave function. 
One can therefore argue that all terms of Eq.(\ref{int}) except the first one 
do not count and the result (\ref{paird}) is exact.

The function $g^{\rm diag1}$ in Eq.\ (\ref{paird}) is plotted in
Figs.\ \ref{dat1}-\ref{dat3} with dashed lines. 
We see that the places of the 
maxima are well described by this approximation. 
Note here that the approximation $g^{\rm diag1}$ indeed shows the Laughlin exchange correlation hole of $\xi^{2 m_2}$.

In order to obtain analytical results for observables 
we can use all three diagonal elements
$|\Psi^{Q{\rm diag}}|^2= C^2 |\Psi_{1}|^2
\left (|\Psi_{2}|^2|\Psi_{3}|^2+|\tilde \Psi_{2}|^2|\tilde \Psi_{3}|^2+|\tilde{\tilde \Psi}_{2}|^2|\tilde {\tilde \Psi}_{3}|^2\right )$.
The advantage of this approximation is that one can compute the expectation
values of observables analytically and get a fairly good approximation. 

We consider the mean distance of the particles 
\be
&&\langle {\left ({\bf r}_1-{\bf r}_2\right )^2\over 2}\rangle=
\int d^2 \xi d^2r d^2 \eta \,
\xi^2 \left |\Psi^{Q {\rm diag}}\right |^2
\nonumber\\&&=
{3 C^2 l_B^2\over\sqrt{1+\frac{4\omega_o^2}{\omega_B^2}}} \left [ (1+2 n_2+|m_2|)+(1+2 n_3+|m_3|)\right ],
\nonumber\\&&
\label{dist}\ee
as well as the area spanned by the three particles,
\be
&&\langle\frac 1 4 |({\bf r}_2\!-\!{\bf r}_1)\times ({\bf r}_3\!-\!{\bf r}_1)|^2\rangle=\int d^2 \xi d^2r d^2 \eta 
\frac 3 4 |\xi\times \eta|^2 \left |\Psi^{Q {\rm diag}}\right |^2
\nonumber\\&&=
\frac 9 4 { C^2 l_B^4\over1+\frac{4\omega_o^2}{\omega_B^2}} \left [ (1+2 n_2+|m_2|)(1+2 n_3+|m_3|)\right ].
\label{area}
\ee
We see that the mean distance and the mean area are determined by
the quantum numbers $1+2 n_2+|m_2|$ and $1+2 n_3+|m_3|$. The mean distance
is given by their sum while the mean area is given by their product.
The nondiagonal terms give smaller and smaller contributions. In table
\ref{tab} we compare the analytical approximate results (\ref{dist}) and
(\ref{area}) with the numerical value of the full wave function. The
agreement especially for the mean distance suggests that this is quite a
good approximation.

\parbox[h]{8cm}{
\begin{table}[h]
\begin{tabular}[]{|l||c|c|c|}
\hline
$n_2m_2n_3m_3$& $1/C$& dist. & area\\
\hline
1100&2.24&2.49&1.13\\
&3&2.5&1.5\\
1111&3.37&4.00&2.25\\
&3&4&6\\
1121&1.83&5.00&7.35\\
&3&5&9\\
1122&4.81&5.49&11.03\\
&3&5.5&10.5\\
2100&2.79&3.50&2.11\\
&3&3.5&2.25\\
2110&1.12&4.49&8.07\\
&3&4.5&6.75\\
2111&1.83&5.00&7.35\\
&3&5&9\\
2120&3.22&5.48&5.07\\
&3&5.5&11.25\\
2121&2.67&5.99&9.94\\
&3&6&13.5\\
2122&3.10&6.46&25.10\\
&3&6.5&15.75\\
2130&2.17&6.44&16.12\\
&3&6.5&15.75\\
2131&3.26&6.92&10.88\\
&3&7&18\\
2132&3.28&7.26&28.59\\
&3&7.5&20.25\\ 
2133&2.27&7.68&14.47\\
&3&8&22.5\\
3110&1.96&5.48&9.90\\
&3&5.5&9\\
3310&2.59&6.46&12.41\\
&3&6.5&11.25\\
3111&1.55&5.99&14.72\\
&3&6&12\\
3311&1.83&6.98&17.45\\
&3&7&15\\
3130&2.58&7.36&16.19\\
&3&7.5&21\\
3330&3.90&8.29&26.75\\
&3&8.5&26.25\\
3131&2.20&7.65&23.02\\
&3&8&24\\
3331&2.40&8.46&23.18\\
&3&9&30\\
3132&3.92&8.18&30.50\\
&3&8.5&27\\
3332&2.28&8.64&32.36\\
&3&9.5&33.75\\
3133&2.37&8.15&23.10\\
&3&9&30\\
3333&3.70&9.29&41.88\\
&3&10&37.5\\
\hline
\end{tabular}
\caption{The normalization constant $C$, the dimensionless mean distance and the area  
for different quantum numbers. The corresponding second lines give the analytical values 
(\protect\ref{dist}) and (\protect\ref{area}) respectively. }\label{tab}
\end{table}
}

\begin{figure}
\psfig{file=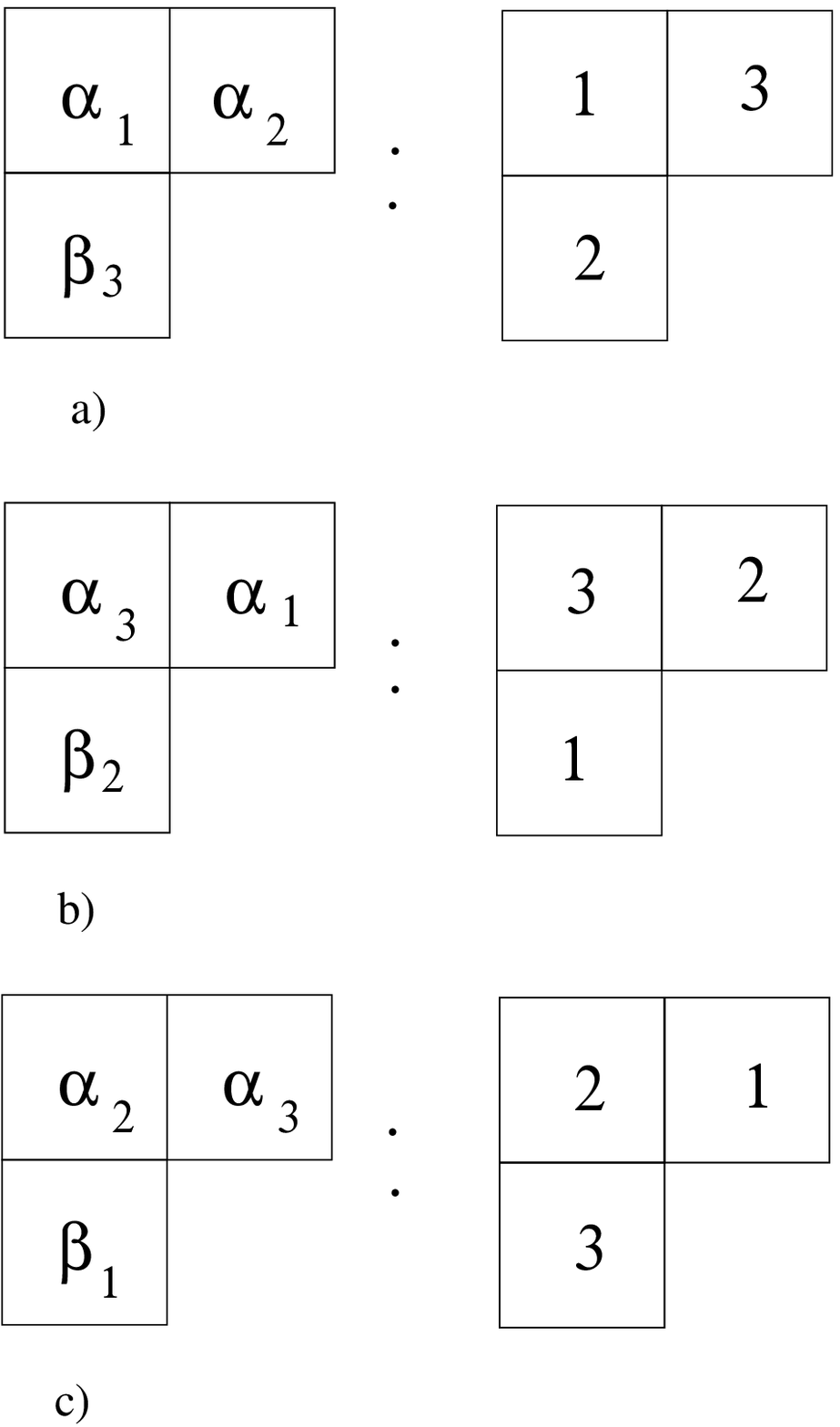,width=8cm}
\caption{The particular Young diagrams contributing to the
spin state $|S=1/2, s_z = 1/2\rangle $ are shown on the left with
the corresponding contributions to the spatial wave function on the
right. The indices in one row are understood to be symmetrized and the
indices in one column are antisymmetrized. 
The diagrams for $|S=1/2, s_z= - 1/2 \rangle $ are obtained
by interchanging $\alpha \leftrightarrow \beta$.}
\label{young}
\end{figure}

Although the states in high magnetic fields are characterized by spin
alignment and are therefore described by the quartet states, we now give
for completeness the other possible state, the doublet state.
The wave functions corresponding to  the doublet states can be
constructed according to  Young's scheme \cite{LL76}.  For the state
$|S=1/2,s_z =1/2 \rangle $  it is given in
Fig.\ \ref{young}. The expressions corresponding to the configurations
(a), (b), and (c) in Fig.\ \ref{young} are obtained by antisymmetrizing the
spin wave function with
respect to the row indices and symmetrizing with respect to
the column ones,
\begin{eqnarray}
&&|1/2, 1/2 \rangle_a = c\left (
\alpha_2\{\alpha_1 \beta_3 -
\alpha_3 \beta_1 \}+\alpha_1\{\alpha_2 \beta_3 -
\alpha_3 \beta_2 \}\right )
\nonumber\\&&
\label{a}
\end{eqnarray}
and correspondingly for (b) and (c).
It is possible to see that one particular spin wave function
transforms into the other one when two particle
indices are interchanged, forming a group. 

The spatial wave functions for each spin diagram are obtained by
transposition of the Young tableaux of
Fig.\ \ref{young}. Once again, the elements in a row must be symmetrized
and the elements in a column antisymmetrized. The spatial wave function
corresponding to (a) in Fig.\ \ref{young}, Eq.\ (\ref{a}), reads
\be
&&\Psi^D_a({\bf r_1},{\bf r_2},{\bf r_3})= C \Psi_{n_1,m_1}(r;\omega_0=0)
\nonumber\\
&&\times
\left \{
\Psi_{n_2,m_2}(-\xi)\Psi_{n_3,m_3}(\eta)
+\Psi_{n_2,m_2}(\tilde \xi)\Psi_{n_3,m_3}(\tilde \eta)
\right .
\nonumber\\&&
\left . 
-\Psi_{n_2,m_2}(\xi) \Psi_{n_3,m_3}(\eta)
-\Psi_{n_2,m_2}(-\tilde{\tilde \xi}) \Psi_{n_3,m_3}(\tilde{\tilde \eta})
\right \}.
\label{Ps}
\ee
Denoting only the spatial indices, the coordinates $\xi,\eta$ defined in
Eq.\ (\ref{coor}) can be abbreviated as $(1,23)$. The coordinate
$\tilde \xi,\tilde \eta$ of (\ref{not}) would correspond to (2,31),
$\tilde {\tilde \xi},\tilde {\tilde \eta}$ to (3,12), and
$-\tilde \xi,\tilde \eta$ to (3,21) respectively. In this short
notation the three orbital wave functions of figure (\ref{young}) read
\be
\Psi_a^D&=&(2,13)+(2,31)-(1,23)-(1,32)
\nonumber\\
\Psi_b^D&=&(1,32)+(1,23)-(3,21)-(3,12)
\nonumber\\
\Psi_c^D&=&(3,21)+(3,12)-(2,31)-(2,13).
\ee

Multiplying the spin and spatial components and
adding all three particular expressions, we get the final wave
function for the doublet state. The final result depends on
the parity of $m_2$. For $m_2$ odd it reads
\begin{eqnarray}
&&\Psi^D({\bf r_1},{\bf r_2}, {\bf r_3})
= C_1 \Psi_{n_1,m_1}(r; \omega_0=0)\nonumber\\
&&\times \big \{
(\beta_1 \alpha_2\alpha_3+\alpha_1 \beta_2 \alpha_3 - 2 \alpha_1 \alpha_2 \beta_3) 
\Psi_{n_2,m_2}(\xi)\Psi_{n_3,m_3}(\eta)
\nonumber\\
&&+(\alpha_1\beta_2\alpha_3+\alpha_1 \alpha_2 \beta_3 - 2 \beta_1 \alpha_2 \alpha_3)
\Psi_{n_2,m_2}(\tilde \xi)
\Psi_{n_3,m_3}(\tilde \eta)
\nonumber\\
&&+(\beta_1\alpha_2\alpha_3+\alpha_1 \alpha_2 \beta_3 - 2 \alpha_1 \beta_2 \alpha_3)
\Psi_{n_2,m_2} (\tilde {\tilde \xi}) \Psi_{n_3,m_3}(\tilde {\tilde \eta})
\big\}
\end{eqnarray}
while for $m_2$ even the wave function is given by 
\begin{eqnarray}
&&\Psi^D({\bf r_1},{\bf r_2}, {\bf r_3})= 
C_2 \Psi_{n_1,m_1}(r; \omega_0=0)
\nonumber\\
&&\times 
\big \{(\alpha_1 \beta_2 \alpha_3 - \beta_1 \alpha_2 \alpha_3) 
\Psi_{n_2,m_2} (\xi)) \Psi_{n_3,m_3}(\eta)
\nonumber\\
&&+
( \alpha_1 \alpha_2 \beta_3 - \alpha_1 \beta_2 \alpha_3)\Psi_{n_2,m_2} ({\tilde \xi}) 
\Psi_{n_3,m_3}({\tilde \eta})
\nonumber\\
&&+
(\beta_1 \alpha_2 \alpha_3 - \alpha_1 \alpha_2 \beta_3)
\Psi_{n_2,m_2} (\tilde {\tilde \xi}) \Psi_{n_3,m_3}(\tilde {\tilde \eta})
\big\}.
\nonumber\\&&
\end{eqnarray}

It is easy to check that the expression for the probability
distribution $|\Psi^D|^2$ in the doublet state
differs from the one for the quartet state by the coefficients in
front of the crossed terms. Our estimate shows
that the character of quantization of the two-particle correlator,
the cluster area, mean distance for doublet states
is qualitatively similar to that for quartet states. Since the
contributions of the crossed terms in the quartet
states are smaller than those in the doublet states, the maxima
in the pair correlation function are more pronounced in the former case.

\section{Conclusions}

The investigation of the three-particle problem in the presence of
magnetic fields is essential to understand the phenomenon of the 
fractional quantum Hall effect \cite{L87}. 
In this paper we have solved exactly the problem of three identical
particles interacting via a spring potential in 2D space in the presence
of an external magnetic field,neglecting the Coulomb interactions. 
We have shown that the wave function
acquires the Jastrow prefactor even in the case of an attractive potential
as a consequence of the anti-symmetric character of the total wave function.
Calculations of the pair correlation function, the area
of the three-particle cluster and the mean distance show a quantization
of these parameters. Particularly, the pair correlation function exhibits
maxima and minima controlled by a quantum number. 

We have also found that the exchange correlation hole at small distances 
shows the behavior known from the Wigner lattices rather than the Laughlin
behavior if we start from the antisymmetrized wave function. However, neglecting 
certain terms based on the physical consideration that the antisymmetrization of
the wave function fails when two particles are fixed for the pair correlation, we obtain
exactly the Laughlin exchange correlation hole. The  corresponding
filling factor is identified with a quantum number appearing in the
exact solution of the model solved here. 

\appendix
\section{Small distance expansion of the paircorrelation function}\label{abc}

The expansion of the paircorrelation function (\ref{pair}) for small $\xi$
can be performed straightforwardly with the help of MATHEMATICA.
After trivial integrations about $d^2r$ we have to calculate 
\be
g(|\xi^*|)&=&\int d^2 \eta^* \left ( 
|\Psi_2|^2|\Psi_3|^2 +2 |{\tilde \Psi_2}|^2|{\tilde \Psi_3}|^2
\right .\nonumber\\
&&\left .
+2 {\rm Re} 
\left \{
2 \Psi_2\Psi_3{\tilde \Psi_2}^*{\tilde \Psi_3}^*+2 {\tilde \Psi_2}{\tilde \Psi_3}{\tilde{\tilde \Psi}_2}^*{\tilde{\tilde \Psi}_3}^*
\right \}
\right )
\label{int}
\ee
where we have used symmetries in the variables and abbreviated $\tilde\Psi_2=\Psi_{n_2,m_2}(\tilde \xi)$, $\tilde{\tilde \Psi}_3=\Psi_{n_3,m_3}(\tilde {\tilde \eta})$ and similar for the other combinations.
Therefore we express the occurring variables (\ref{not}) in the wave function in terms of the integration variables 
\be
|\tilde \xi|^2&=&\frac 1 4 \left [3 |\eta|^2+|\xi|^2-2 \sqrt{3} |\eta||\xi| \cos{(\phi_\xi-\phi_\eta)}\right ];
\nonumber\\
{\rm e}^{i\phi_{\tilde \xi}}&=&{1 \over 2 |\tilde\xi|} \left (\sqrt{3} |\eta| {\rm e}^{i \phi_\eta}-|\xi|{\rm e}^{i \phi_\xi} \right )
\label{rule}
\ee
and similarly for $\tilde \eta$ as well as $\tilde{\tilde \eta}$ and $\tilde{\tilde \xi}$. Using the dimensionless coordinates (\ref{diff}) in the wave function (\ref{wave}) and changing variables (\ref{rule}), the integral (\ref{int}) is quite massy. However, expanding in small $\xi^*$ the angular dependence $\phi_\eta-\phi_\xi$ drops out and we obtain the form (\ref{abc1})
\be
g(|\xi^*|)&=&a |\xi^*|^{2 m_2}+(b+c |\xi^*|^{2 m_2})|\xi^*|^2 +o(|\xi^*|^3)
\ee
where the remaining modulus integral $d |\eta|$ can be performed for the constants $a$ and $c$ with the result
\be
a&=&2 {(n_2+m_2)!\over m_2!^2 n_2!}
\nonumber\\
c&=&-{2(n_2+m_2)!\over n_2! m_2!} \left (2 {n_2\over (1+m_2)!}+{1\over m_2!}
\right ).
\ee
Therefore we see that the constant $a$ is not vanishing which means that the Laughlin liquid behavior is not resumed if full antisymmetrization is pertained for the pair correlation function.

For completeness let us also give the constant $b$ which integral can be done
only numerically [$x=|\eta^*|^2$]
\be
b&=&
    {(-1)^{2m_3}3^{m_2-1}\over 4^{m_2+m_3}}
{n_2!n_3!\over
    ( m_2 + n_2 ) !
    (m_3 + n_3) !}
 {\cal I}
\nonumber\\
{\cal I}&=&\int\limits_0^\infty d x x^{m_2+m_3-1}{\rm e}^{-x}
\nonumber\\
&&\times\left \{8(m_2-3 m_3)^2 L_{n_2}^{m_2}({3x \over 4})^2L_{n_3}^{m_3}({x\over 4})^2\right .\nonumber\\
&&\left .+12 (m_2-3 m_3) x L_{n_2}^{m_2}({3 x\over 4})L_{n_3}^{m_3}({x\over 4})
\right .\nonumber\\
&&\left .\,\,\,\times
\left (
L_{n_2}^{m_2}({3 x\over 4})L_{n_3-1}^{m_3+1}({x\over 4})-L_{n_2-1}^{m_2+1}({3 x\over 4})L_{n_3}^{m_3}({x\over 4})
\right )
\right .\nonumber\\
&&\left .+9 x^2 \left ( L_{n_2}^{m_2}({3 x\over 4})L_{n_3-1}^{m_3+1}({x\over 4})-L_{n_2-1}^{m_2+1}({3 x\over 4})L_{n_3}^{m_3}({x\over 4})\right )^2
\right \}
\nonumber\\&&
\ee  

\end{document}